\documentclass[prb,twocolumn]{revtex4-1}
\usepackage{amsmath}
\usepackage{pxfonts}
\usepackage{amsfonts}
\usepackage{amssymb}
\usepackage{graphicx}

\begin{document}

\title{Transport of Indirect Excitons in a Potential Energy Gradient}

\author{J.R.~Leonard} \email{jleonard@physics.ucsd.edu} \author{M.~Remeika} \author{M.K.~Chu} \author{Y.Y.~Kuznetsova} \author{A.A.~High} \author{L.V.~Butov} \affiliation{Department of Physics, University of California at San Diego, La Jolla, CA 92093-0319, USA}

\author{J.~Wilkes} \affiliation{Department of Physics and Astronomy, Cardiff University, Cardiff CF24 3AA, United Kingdom}

\author{M.~Hanson} \author{A.C.~Gossard}
\affiliation{Materials Department, University of California at Santa Barbara, Santa Barbara, CA 93106-5050, USA}

\begin{abstract}
We realized a potential energy gradient - a ramp - for indirect excitons using a shaped electrode at constant voltage. We studied transport of indirect excitons along the ramp and observed that the exciton transport distance increases with increasing density and temperature.
\end{abstract}

\maketitle

An indirect exciton in a coupled quantum well structure (CQW) is a bound state of an electron and a hole in separate wells (Fig. 1a). The spatial separation allows one to control the overlap of electron and hole wavefunctions and engineer structures with lifetimes of indirect excitons orders of magnitude longer than those of direct excitons. Long lifetimes of the indirect excitons allow them to travel over large distances before recombination \cite{Hagn95, Butov98, Larionov00, Butov02, Voros05, Ivanov06, Gartner06, Hammack09, Dubin11, Dubin12}. Furthermore, indirect excitons have a built-in dipole moment $ed$, where $d$ is close to the distance between the quantum well (QW) centers that allows their energy to be controlled by voltage: an electric field $F_z$ perpendicular to the QW plane results in the exciton energy shift $E = e d F_z$ \cite{Miller85}. These properties allow studying transport of indirect excitons in electrostatically created in-plane potential landscapes $E(x,y) = e d F_z(x,y)$.

Exciton transport was studied in various electrostatic potential landscapes including circuit devices \cite{High07, High08, Grosso09}, traps \cite{High09}, lattices \cite{Remeika09, Remeika12}, moving lattices - conveyers - created by a set of ac voltages \cite{Winbow11}, and narrow channels \cite{Vogele09, Grosso09, Cohen11}. Several exciton transport phenomena have been observed, including the inner ring in emission patterns \cite{Butov02, Ivanov06, Hammack09, Dubin12, Stern08, Ivanov10}, transistor effect for excitons \cite{High07, High08, Grosso09}, localization-delocalization transition in random potentials \cite{Butov02, Ivanov06, Hammack09} and in lattices \cite{Remeika09, Remeika12}, and dynamical localization-delocalization transition in conveyers \cite{Winbow11}. Exciton transport was also studied in potential energy gradients created by voltage gradients in electrodes \cite{Hagn95, Gartner06}.

In this work, we study exciton transport in a potential energy gradient - a ramp - created by a shaped electrode at constant voltage. We utilize the ability to control exciton energy by electrode shape \cite{Kuznetsova10} and design the shape of a top electrode on the sample surface so that a voltage applied to it creates a constant potential energy gradient for indirect excitons in the CQW. The excitonic ramp realizes directed transport of excitons as a diode realizes directed transport of electrons.

The advantages of this shaped-electrode-method include the suppression of heating by electric currents in electrodes (such currents may appear in the case when the ramp potential is created by a voltage gradient in the top electrode) and the opportunity to engineer the exciton energy profile along the ramp by designing the electrode shape. We also measure exciton transport in a narrow channel formed by a voltage applied to an electrode stripe of constant width - a flat-energy channel \cite{Grosso09, Cohen11}.

The CQW structure is grown by molecular beam epitaxy. An $n^+$-GaAs layer with $n_{Si}$ = $10^{18}$ cm$^3$ serves as a homogeneous bottom electrode. A semitransparent top electrode is fabricated by depositing a 100 nm indium tin oxide layer. Two 8 nm GaAs QWs separated by a 4 nm Al$_{0.33}$Ga$_{0.67}$As barrier are positioned 100 nm above the $n^+$-GaAs layer within an undoped $1\,\mu$m thick Al$_{0.33}$Ga$_{0.67}$As layer. Positioning the CQW closer to the homogeneous electrode suppresses the in-plane electric field \cite{Hammack06}, which otherwise can lead to exciton dissociation \cite{Zimmerman97}. Excitons are photoexcited by a 633 nm HeNe laser focused to a spot with full width half maximum $4\,\mu$m.

\begin{figure}[htbp]
\centering
\includegraphics[width=7.5cm]{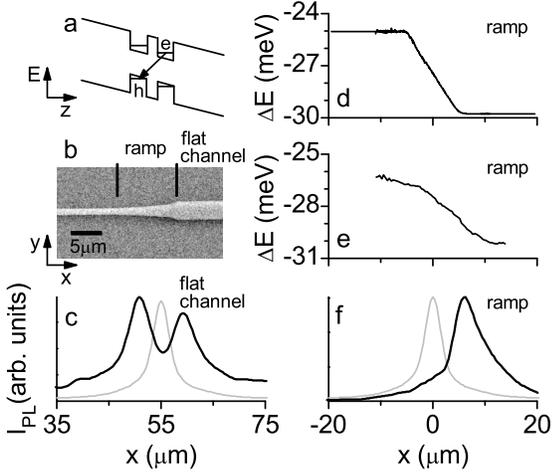}
\caption{(a) CQW band diagram. e, electron; h, hole. (b) Scanning electron microscope image of the electrode, which forms the linear potential energy gradient - ramp (center) and flat-energy channels (sides) for indirect excitons. (c) Indirect exciton PL profile in the flat channel (black). $P_{ex} = 0.5$ $\mu$W, $T_{bath} = 1.6$ K. Laser excitation profile is shown in gray. (d) Simulated potential energy for indirect excitons in the ramp surrounded by flat-energy channels (thin lines show extrapolations). (e) Indirect exciton PL energy in the ramp. $P_{ex} = 0.02$ $\mu$W, $T_{bath} = 5.4$ K. (f) Indirect exciton PL profile in the ramp (black). $P_{ex} = 0.5$ $\mu$W, $T_{bath} = 1.6$ K. Laser excitation profile is shown in gray. $V_e = -3$ V for all data.}
\end{figure}

Figure 1c shows the profile of the indirect exciton photoluminescence (PL) intensity $I(x)$ in the $3\,\mu$m channel. The laser excitation profile is shown in gray. The emission pattern has two maxima around the excitation spot and is nearly symmetric relative to the excitation spot position. This pattern is similar to the inner ring  studied previously \cite{Butov02, Ivanov06, Hammack09, Dubin12, Stern08, Ivanov10}. The inner ring was explained in terms of exciton transport and cooling: optical excitation heats the exciton gas, excitons cool towards the lattice temperature as they travel away from the excitation spot, the cooling results in an increase in the occupation of the low-energy optically active exciton states and, as a result, the appearance of an emission ring around the excitation spot \cite{Butov02, Ivanov06, Hammack09, Dubin12, Ivanov10}.

Figure 1d presents a numerical simulation of the potential energy profile for indirect excitons $e d F_z$ for constant voltage $V_e = -3$ V applied to the top electrode shown in Fig. 1b. The principle of the ramp is the following: a thinner electrode produces a smaller $F_z$ due to field divergence near the electrode edges, therefore narrowing the electrode increases the exciton energy thus creating a ramp potential for excitons. The electrode shape (Fig. 1b) was designed to obtain a linear ramp with a constant gradient of the exciton energy (Fig. 1d). The energy gradient for indirect excitons in the ramp can be controlled by voltage - it is proportional to $V_e$. The ramp is surrounded by flat-energy channels where the electrode width is constant ($1\,\mu$m left of the ramp and $3\,\mu$m right of the ramp). The energy of indirect excitons is constant in the flat-energy channels (Fig. 1d).

Figure 1e shows the measured spatial profile of the emission energy of indirect excitons along the ramp $E(x) = \int E I(x,E) dE / I(x)$, where $I(x) = \int I(x,E) dE$. The measurement was performed at $T_{bath} = 5.4$ K and $P_{ex} = 0.02 \mu$W: a higher temperature was chosen to reduce the effect of QW disorder and a low laser excitation - to reduce the effect of exciton-exciton repulsion, these effects are described below. The energy of the indirect excitons was measured relative to the direct exciton energy, which is practically independent on $x$. The measured (Fig. 1e) and calculated (Fig. 1d) ramp profiles are qualitatively similar (note that the measured energy difference between the direct and indirect exciton also includes the difference between the direct and indirect exciton binding energies, a few meV for the CQW \cite{Butov99}).

Figure 1f shows the profile of the indirect exciton PL intensity $I(x)$ in the ramp. The laser excitation profile is shown in gray. In contrast to the flat-energy channel (Fig. 1c), on the ramp (Fig. 1f), $I(x)$ is asymmetric relative to the excitation spot due to exciton transport along the potential energy gradient.

\begin{figure}[htbp]
\centering
\includegraphics[width=5.5cm]{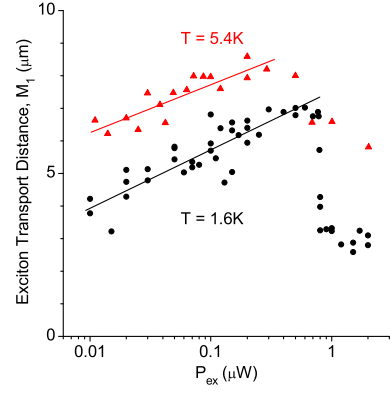}
\caption{The average transport distance of indirect excitons along the ramp $M_1$ as a function of excitation density $P_{ex}$ for $T_{bath} = 1.6$ (black circles) and 5.4 K (red triangles), $V_e = -3$ V.}
\end{figure}

We studied exciton transport in the ramp for different excitation powers $P_{ex}$ and cryostat temperatures $T_{bath}$. Exciton transport in the ramp is presented by the extension of the exciton cloud along the energy gradient (Fig. 2). We quantify it by the first moment of the PL intensity $M_1 = \int x I(x) dx/ \int I(x) dx$, which characterizes the average transport distance of indirect excitons along the ramp. Figure 2 shows that the exciton transport distance along the ramp increases with density. We attribute this to the screening of the disorder in the structure by indirect excitons. The screening originates from the repulsive interaction between indirect excitons, which are dipoles oriented perpendicular to the QW plane \cite{Ivanov02, Ivanov10}. Such screening improves the exciton mobility and, as a result, increases their transport distance along the energy gradient. Figure 2 also shows that exciton transport distance along the ramp increases with temperature. We attribute this to the thermal activation of indirect excitons in the disorder potential that facilitates their transport. These interpretations are in agreement with the theoretical model presented below. A drop of $M_1$ observed at the highest excitation powers can be related to a photoexcitation-induced reduction of the electric field $F_z$ in the device.

The measurements of exciton transport in a ramp and in a flat-energy region allow estimating the exciton temperature $T$. The former measurement provides an estimate for the exciton mobility $\mu_{\rm x}$, the latter - for the exciton diffusion coefficient $D_{\rm x}$, and the Einstein relationship between them - for the exciton temperature. The exciton diffusion coefficient can be estimated using $D_{\rm x} \approx w^2 / \tau_{\rm opt}$, where $w$ is the half-width at half-maximum of the indirect exciton emission cloud in the flat-energy region and $\tau_{\rm opt}$ is the indirect exciton lifetime ($\tau_{\rm opt} \approx 50$ ns in the structure). The exciton mobility can be estimated using $\mu_{\rm x} = -\frac{M_1}{\tau_{\rm opt} \nabla U_{ramp}}$, where $\nabla U_{ramp}$ is the exciton potential energy gradient in the ramp. For $T_{bath} = 1.6$ K and $P_{ex} = 0.02$ $\mu$W, the estimate gives $D_{\rm x} \sim 3$ cm$^{2}$/s, $\mu_{\rm x} \sim 3000$ cm$^2$/eVs, and $T = \frac{D_{\rm x}}{k_B \mu_{\rm x}} \sim 10$ K ($k_B$ is the Boltzmann constant). This estimate is consistent with the exciton temperature in the excitation spot obtained within the theoretical model for the exciton energy relaxation and transport in a ramp (Fig. 3a). This model is described below.

We modeled the in plane transport of excitons in the ramp using the following non-linear partial differential equation for exciton density $n_{\rm x}$:
\begin{equation}
\nabla\left[D_{\rm x} \nabla n_{\rm x} + \mu_{\rm x} n_{\rm x} \nabla\left(u_0n_{\rm x} + U_{ramp}\right)\right] - n_{\rm x}/\tau_{\rm opt} + \Lambda = 0.
\label{a}
\end{equation}
The first and second terms in the square brackets account for exciton diffusion and drift, respectively. In the one dimensional geometry relevant to the experiment, $\nabla=\partial/\partial x$. The exciton dipole-dipole repulsion, which is approximated by $u_0 n_{\rm x}$ with
$u_0 = 4 \pi d e^2 / \varepsilon_b$ ($\varepsilon_b$ is the background dielectric constant), \cite{Ivanov02, Ivanov06, Ivanov10} and the ramp potential $U_{ramp} = e d F_z$ are included in the drift term. The decay rate of excitons is given by $1/\tau_{\rm opt}$. $\Lambda(x)$ is the exciton generation rate. The diffusion coefficient and mobility are related by the generalized Einstein relationship $\mu_{\rm x} = D_{\rm x} / k_B T_0 \left(e^{T_0/T} - 1\right)$, where $T_0 = \left(2\pi\hbar^2n_{\rm x}\right) / \left(M_{\rm x}gk_B\right)$ is the quantum degeneracy temperature. Here, $g = 4$ is the spin degeneracy. The diffusion coefficient is given by the thermionic model $D_{\rm x} = D_0 e^{-U_0/\left(u_0n_{\rm x} + k_BT\right)}$, where $U_0/2 = 0.75$ meV is the amplitude of disorder potential in the structure \cite{Ivanov02}. In addition, we included a thermalization equation to account for energy relaxation of photoexcited excitons:
\begin{equation}
S_{phonon}\left(T_0, T\right) = S_{pump}\left(T_0,T,\Lambda,E_{inc}\right),
\label{thermalization}
\end{equation}
where $S_{phonon}$ is the cooling rate due to bulk longitudinal acoustic phonon emission, $S_{pump}$ is the heating rate due to the laser excitation and $E_{inc}$ is the excess energy of photoexcited excitons (17 meV in the simulations). Expressions for $S_{phonon}$, $S_{pump}$, $\tau_{\rm opt}$ and all other parameters of the model are given in \cite{Ivanov02, Ivanov06, Hammack09}.

\begin{figure}[htbp]
\centering
\includegraphics[width=5.5cm]{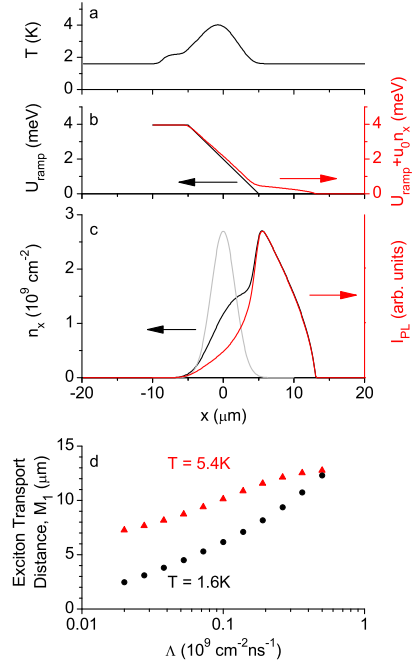}
\caption{Theoretical simulations. (a) Exciton temperature. (b) Bare ramp potential $U_{ramp}$ and screened ramp potential $U_{ramp} + u_0n_{\rm x}$. (c) Exciton density (black) and PL intensity (red). Laser excitation profile is shown in gray. $\Lambda = 10^8$ cm$^{-2}$ ns$^{-1}$, $T_{bath} = 1.6$ K. (d) The average transport distance of indirect excitons along the ramp $M_1$ as a function of exciton generation rate $\Lambda$ for $T_{bath} = 1.6$ (black) and 5.4 K (red).}
\end{figure}

The results of the simulations are presented in Fig. 3. The calculated exciton temperature in the excitation spot (Fig. 3a) is close to the temperature estimate above. Figure 3b shows the bare ramp potential $U_{ramp}$ and the ramp potential screened by the exciton-exciton repulsion $U_{ramp} + u_0n_{\rm x}$. Figure 3c shows the density and PL intensity of indirect excitons. The latter is qualitatively similar to the measured PL intensity (Fig. 1f). Figure 3d shows the average transport distance of indirect excitons along the ramp $M_1$ as a function of exciton generation rate $\Lambda(x)$. Within the model, the exciton transport distance along the ramp increases with density due to the screening of the disorder in the structure by repulsively interacting indirect excitons and increases with temperature due to the thermal activation of indirect excitons in the disorder potential. These features observed in the model are qualitatively similar to the experimental data, compare Figs. 2 and 3d.

In summary, we report on the realization of an electrostatic ramp potential for indirect excitons using a shaped electrode at constant voltage and on experimental and theoretical studies of exciton transport along the ramp.

This work was supported by NSF Grant No. 0907349. J.W. was supported by EPSRC, A.A.H. was supported by an Intel fellowship.

\end{document}